\documentclass{aa}
\usepackage[dvips]{graphics}

\begin{document}
 
   \thesaurus{07     
              (13.09.6;  
               08.12.1;  
               08.01.3;  
               08.03.4;  
               08.22.3;  
               04.19.1)} 
   \title{Water vapor absorption in early M-type stars}
 
 
   \author{M.~Matsuura\inst{1, 2},
           I.~Yamamura\inst{3, 2}, 
           H.~Murakami\inst{1}, 
           M.M.~Freund\inst{1, 4}, 
           M.~Tanaka\inst{1}}
 
   \offprints{M.M.
              mikako@astro.isas.ac.jp}
 
   \institute{Institute of Space and Astronautical Science (ISAS),
              3-1-1 Yoshinodai, Sagamihara, Kanagawa 229-8510, Japan
         \and
              Department of Astronomy, School of Science, 
              University of Tokyo, 7-3-1 Hongo, Bunkyo, Tokyo 113-0033, Japan
         \and
              Astronomical Institute `Anton Pannekoek',
              University of Amsterdam, 
              Kruislaan 403, 1098 SJ, Amsterdam, The Netherlands
         \and
              NASA Ames Research Center, MS 239-4, Moffett Field, 
              CA 94035-1000, USA}

   \authorrunning{M. Matsuura et al.}
   \titlerunning{Water vapor absorption in early M-type stars}
 
   \date{Received ----; accepted  ----}
 
   \maketitle
 
   \begin{abstract}

 The spectrometers onboard the Infrared Telescope in Space (IRTS)
reveal water vapor absorption in early M-type stars, as early as M2.
Previous observations detected H$_2$O vapor absorption
only in stars later than M6, with the exception of the recent
detection of H$_2$O in \object{$\beta$~Peg} (M2.5 II-III).
In our sample of 108 stars, 67 stars have spectral types earlier than
M6. The spectral types are established by means of their near-infrared
colors on a statistical basis. 
Among the 67 stars of spectral types earlier than M6, we find water vapor
absorption in six stars.
The observed absorption features are interpreted using
a local thermodynamic equilibrium model.
The features are reasonably fitted by model spectra with
excitation temperatures of 1000--1500~K
and water column densities of $5\times10^{19}$ to $1\times10^{20}$~cm$^{-2}$.
These numbers imply that the H$_2$O molecules are present 
in a region of the atmosphere, located above the photosphere.
Furthermore, our analysis shows a good correlation between
the H$_2$O absorption band strength,
and the mid-infrared excess due to the circumstellar dust. 
We discuss the relation between the outer atmosphere and the mass loss.

   \end{abstract}
 
%
 
\section{Introduction}
 
 Water is one of the most abundant molecules in the atmosphere of late
M-giants, and it is a dominant absorber in the near-infrared (near-IR) region. 
Water vapor in stellar atmospheres has been studied
by theoretical and observational methods.
The strength of the H$_2$O absorption possibly correlates with
the spectral type, the effective temperature of the star, and
the near-IR color 
(Kleinmann \& Hall \cite{kleinmann}; 
Lan\c{c}on \& Rocca-Volmerange \cite{lancon}), 
even though such correlations were not clearly found by Hyland (\cite{hyland}). 
According to Spinrad \& Wing (\cite{spinrad}) and Hyland (\cite{hyland}),
H$_2$O could only be detected in giants with spectral type M6 or later.
This was
consistent with hydrostatic models of the atmosphere of red-giants
(Tsuji \cite{tsuji}; Scargle \& Strecker \cite{scargle}). 
As a further complication, most late M-giants are long period variables, and 
in general, the band strength of the H$_2$O absorption
features depends on stellar variability.
For example, Mira variables show very deep H$_2$O absorption,
and the depths of H$_2$O features in Miras vary
from phase to phase (Hyland \cite{hyland}). 
High-resolution spectroscopic observations of the Mira
variable R~Leo by Hinkle \& Barnes (\cite{hinkle}) revealed that a
significant fraction of the H$_2$O molecules were in a component with a
distinct velocity, and a cooler excitation temperature than molecules near
the photosphere. They interpreted this `cool component' as an overlying layer
above the photosphere.

 These previous studies were mostly based on ground
or airborne observations, 
where the terrestrial H$_2$O interferes with a detailed study 
of the center of the stellar water bands.
In contrast, observations from space are
ideal for investigations of stellar H$_2$O features. 
Using the Infrared Space Observatory (ISO), 
Tsuji et al. (\cite{tsuji97}) discovered a weak H$_2$O absorption
in the early M-type star, \object{$\beta$~Peg} (M2.5II-III).
They argued that the observed H$_2$O is in a `warm molecular 
layer' above the photosphere.

 In this paper, we present the results of a study of H$_2$O
absorption features, using data from 
the Infrared Telescope in Space 
(IRTS, Murakami et al. \cite{murakami96} and references therein).

\section{Sample Data}
 This study is based on data from the
two grating spectrometers onboard the IRTS:
the Near-Infrared Spectrometer (NIRS), and the 
Mid-Infrared Spectrometer (MIRS). The IRTS was launched in March 1995, and
it surveyed about 7\% of the sky with
four instruments during its 26 day mission. 
The NIRS covers the wavelength region
from 1.43 to 2.54 and from 2.88 to 3.98~$\mu$m in 24 channels 
with a spectral resolution for point sources of $\Delta\lambda = 0.10-0.12$~$\mu$m. 
The MIRS covers the range from 4.5 to 11.7~$\mu$m in 32 channels
with a resolution of $\Delta\lambda = 0.23-0.36~\mu$m. 
Both spectrometers have a rectangular entrance aperture of 
$8\arcmin\times8\arcmin$. 
The total number of detected point sources is about 50,000 
for the NIRS (Freund et al. \cite{freund}) and about 1,000 
for the MIRS (Yamamura et al. \cite{yamamura96}).
The estimated absolute
calibration errors are 5\% for the NIRS, and 10\% for the MIRS.
This may cause systematic errors in the
colors and the H$_2$O index discussed in this paper.
The spectra are not color corrected. 

All stars used in this study were observed both by the NIRS and the MIRS
between April 9 and 24.
We only include stars at high galactic latitudes ($|b|>10\degr$)
in this sample,
to minimize source confusion and interstellar extinction.
Each selected star has a unique association with the IRAS
Point Source Catalog (PSC, \cite{iras-psc}), within 8$\arcmin$
from the nominal NIRS position, and each associated PSC entry has reliable
12 and 25~$\mu$m band fluxes (FQUAL$=$3; IRAS Explanatory Supplement
\cite{iras-es}).
Known carbon and S-type stars
(Stephenson \cite{stephenson89}, \cite{stephenson84}), 
as well as non-stellar objects were discarded from the sample. 
The current sample contains 108 stars.
The signal-to-noise ratio for all stars is larger than 5 in all NIRS bands.

Of the 108 stars used in this study, 43 have a unique association in the 
Bright Star Catalogue (BSC, Hoffleit \& Jaschek \cite{bsc}). 
The distribution of spectral types for these 43 stars are:
1~B, 2~F, 4~G, 22~K, and 14~M-types.
%
%
A total of 31 M-giants are found in the General Catalogue of Variable Stars (GCVS, Kholopov et al.~\cite{gcvs}), and the
distribution of variable types in the GCVS is as follows: 
10 Miras, 13 semi-regulars (SR, SRa or SRb, hereafter SR), and 8 irregulars (L or Lb, hereafter L). 
No M-dwarfs or M-supergiants were found in the BSC or the GCVS associations.

\section{Results}

\begin{figure}
\resizebox{\hsize}{!}{\includegraphics{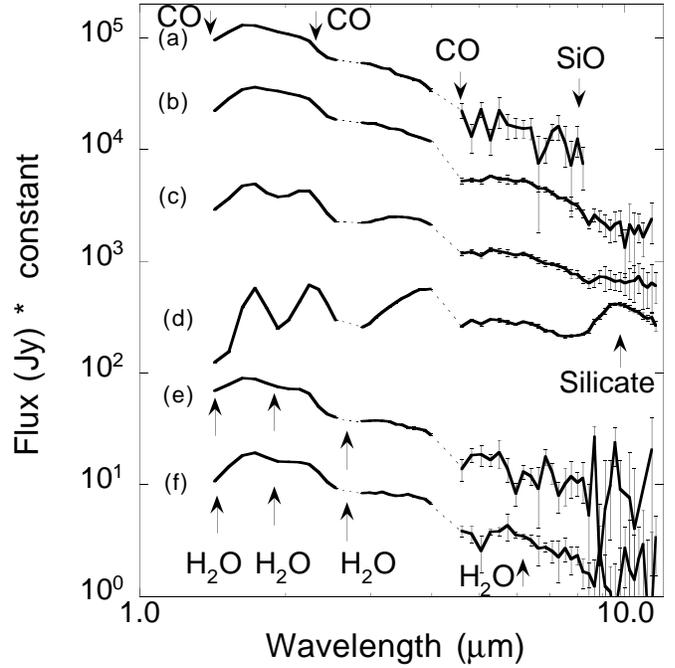}}

\caption{
The combined NIRS \& MIRS spectra of M-giants are shown. 
From top to bottom: (a) \object{HR~1667} (M2III), (b) \object{HR~257} (M4III)
(c) \object{X~Hor} (M6-M8, SRa), (d) \object{RR~Aql} (M6e-M9, Mira), 
(e) \object{AK~Cap} (M2, Lb) and (f) \object{V~Hor} (M5III, SRb).
The error bars represent the noise in the subtracted background level.
Water absorption bands at 1.9 and 2.7~$\mu$m are 
seen in the two early M-giants (e) \object{AK~Cap} and (f) \object{V~Hor},
in contrast to other early M-giants ((a) and (b)).
}
\label{spec}
\end{figure}

In Fig.~\ref{spec}, we show the composite spectra of six representative
M-giants observed by the NIRS and the MIRS.
We indicate the position of the molecular absorption features due to 
CO (1.4, 2.3, 4.6~$\mu$m),
H$_2$O (1.5, 1.9, 2.7, 6.2~$\mu$m), and SiO (8.2~$\mu$m).
The broad-band emission at 9.7~$\mu$m is due to silicate dust.
The H$_2$O bands at 1.9 and 2.7~$\mu$m are
visible in two early M-giants, \object{AK~Cap} (M2, Lb) and
\object{V~Hor} (M5, SRb), where no H$_2$O in the photosphere
was expected to be detectable.

\begin{figure}
\resizebox{\hsize}{!}{\includegraphics{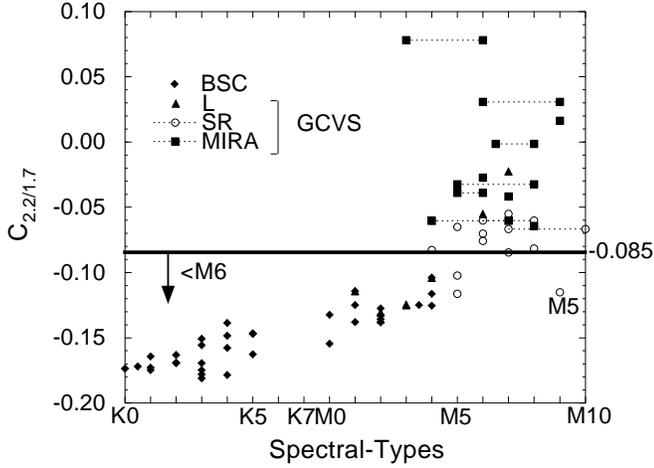}}
\caption{
The color $C_{2.2/1.7}$ is plotted against the spectral type
derived from BSC and GCVS. 
Ranges of spectral types of some SRs and Miras are represented by dotted bars.
The thick horizontal line indicates the boundary of stars with 
spectral types $<$~M6.
}
\label{spec_color}
\end{figure}

One could argue that the early M-type stars with H$_2$O absorption had
spectral types later than M6 at the time of observation, because of their
variability.  However, this is not the case.
We can estimate the spectral types, using the relation between the spectral
type and color (Bessell \& Brett \cite{bessell88}). We use the color
$C_{2.2/1.7}$ instead of the photometric color $H-K$, where $C_{2.2/1.7}$
is defined as:
\begin{equation}
   C_{2.2/1.7} = \log (F_{2.2} / F_{1.7}).
\end{equation}
$F_{2.2}$ and $F_{1.7}$ are the IRTS/NIRS fluxes 
at 2.2 and 1.7~$\mu$m in units of Jy, respectively.
For the NIRS wavelength region, 
the 2.2 and 1.7~$\mu$m bands are least affected by stellar 
H$_2$O absorption bands.
Fig.~\ref{spec_color} shows $C_{2.2/1.7}$ versus the spectral types 
from BSC and GCVS
of all the known K- and M-giants (59 giants) in the sample. 
There is a clear increase in $C_{2.2/1.7}$ toward later spectral type
except for Miras.
One M9 star (\object{KP~Lyr} $=$ \object{ADS 11423}) deviates from this relation,
but we regard this star as M5, according to Abt (\cite{abt}).

All stars of spectral type M6 and later are 
above $C_{2.2/1.7}=-0.085$.
Thus, stars bluer than $-0.085$ are expected to have spectral types
earlier than M6.
There are 67 stars in the $<$~M6 region, defined by 
$C_{2.2/1.7}<-0.085$.
$C_{2.2/1.7}$ for \object{AK~Cap} and \object{V~Hor} is $-0.130$ and $-0.102$,
respectively.
These numbers confirm that these 2 stars have
spectral types earlier than M6 at the time of the IRTS observation,
even though both stars show clear H$_2$O absorption.
%
%
%
%

\begin{figure}
\resizebox{\hsize}{!}{\includegraphics{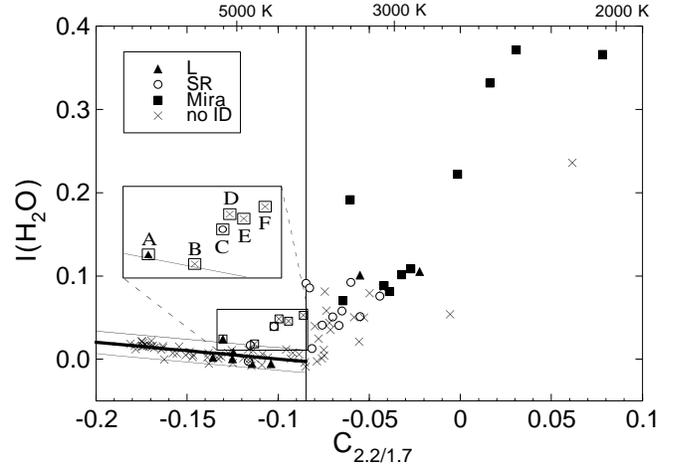}}
\vspace{0cm}
\caption{
$I_\mathrm{H_{2}O}$ is plotted as a function of near-IR color $C_{2.2/1.7}$.
Symbols represent variable types in GCVS.
The blackbody color temperatures corresponding to $C_{2.2/1.7}$ are indicated
on the upper x-axis.
Left of the vertical line at $-0.085$ is the region of stars $<$~M6. 
The thick horizontal line indicates
$I_\mathrm{H_{2}O}$ v.s. $C_{2.2/1.7}$ relation
for stars with spectral type $<$~M6.
Six stars above the upper thin line ($+2\sigma$) with spectral types 
$<$~M6 are supposed to have H$_2$O absorption.
Two stars from the sample lie outside of this plot at
$(C_{2.2/1.7},I_\mathrm{H_{2}O})=(0.888, 0.301), (-0.213, 0.025)$.
}
\label{h2o_nirs}
\end{figure}

We now discuss the relationship between the H$_2$O absorption strength and 
the spectral type. For this, we define the H$_2$O index $I_\mathrm{H_{2}O}$ as follows:
\begin{equation} 
 I_\mathrm{H_{2}O} = \log (F_\mathrm{cont} / F_{1.9}) ,
\end{equation}
where $F_\mathrm{cont}$ is the continuum flux level at 1.9~$\mu$m in units of Jy, 
which is evaluated by linear
interpolation between $F_{1.7}$ and $F_{2.2}$. 
$F_{1.9}$ is the observed flux (Jy) at 1.9~$\mu$m. 
In Fig.~\ref{h2o_nirs}, we plot $I_\mathrm{H_{2}O}$ as a function 
of $C_{2.2/1.7}$.
The dominant measurement errors for $C_{2.2/1.7}$ and $I_\mathrm{H_{2}O}$
are due to by the slitless spectroscopy, and they 
are roughly 0.01 and 0.002 for stars $<$~M6, respectively.

\begin{figure}
\resizebox{\hsize}{!}{\includegraphics{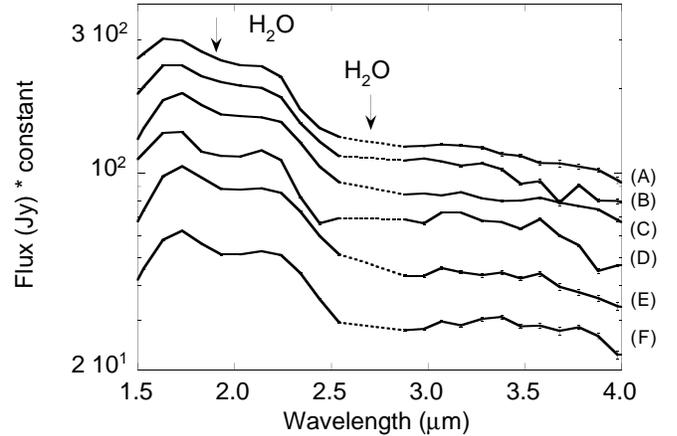}}
\vspace{0cm}
\caption{
The six early M-type stars with clear H$_2$O absorption bands, which are
above the $+2\sigma$ line in Fig.~\ref{h2o_nirs}. From top to
bottom: (A) \object{AK~Cap}, (B) \object{IRAS~21222$-$4155}, (C)
\object{V~Hor}, (D) \object{IRAS~21269$-$3711}, (E)
\object{IRAS~20073$-$1041}, and (F) \object{IRAS~05124$-$4936}.
}
\label{earlym}
\end{figure}

Fig.~\ref{h2o_nirs} can be used to find candidate stars ($<$~M6) with
H$_2$O absorption.
Since we estimated $F_\mathrm{cont}$ by linear interpolation,
$I_\mathrm{H_{2}O}$ is not zero even in the absence of H$_2$O.
We evaluate the relation between $I_\mathrm{H_{2}O}$ and $C_{2.2/1.7}$ 
for stars without H$_2$O, 
using a linear fit for the 67 stars in the $<$~M6 region,
by minimizing a merit function $\sum_{i}^{n}|y_{i}-a-bx_{i}|$ for $n$
data points $(x_{i},y_{i})$ (Press et al. \cite{press}).
This fit is robust against outliers, i.e. stars with H$_2$O. 
The result is indicated as a thick line.
The thin lines indicate $\pm2\sigma$ level from the fit,
where $\sigma$ is standard deviation of $I_\mathrm{H_{2}O}$.
Here, we use the $+2\sigma$ level as a
threshold to find candidate stars with H$_2$O,
because no star appears below the $-2\sigma$ line.
The stars above the $+2\sigma$ are supposed to be stars 
with H$_2$O absorptions.
There are 6 such stars in the region of $<$~M6.
The value of $2\sigma$ is almost equal
to that of 3 standard deviations after the outliers are excluded.
The spectra of the stars are shown in Fig.~\ref{earlym}.
These 6 stars above the $+2\sigma$ line show clear evidence of H$_2$O absorption
bands at 1.9 and 2.7~$\mu$m.
Only \object{AK~Cap} and \object{V~Hor} have
identifications of spectral type in BSC or GCVS.
On the basis of their $C_{2.2/1.7}$ (ranging from $-0.113$ to $-0.086$),
the other four stars are probably M-type stars, and not K-type stars
(see Fig.~\ref{spec_color}).

%
%
%
%

\section{Discussion}

\begin{figure}
\resizebox{\hsize}{!}{\includegraphics{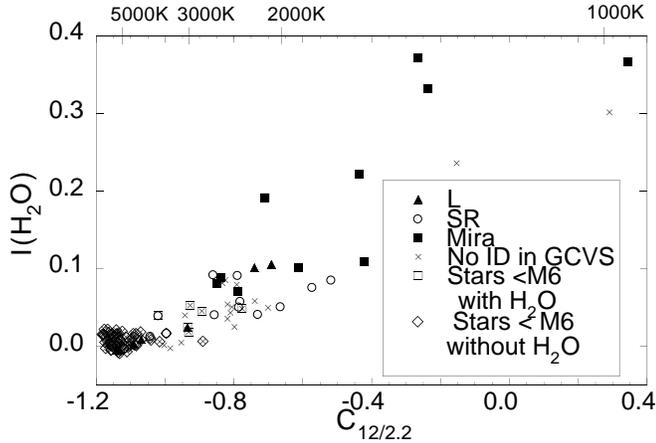}}
\vspace{0cm}
\caption{
$I_\mathrm{H_{2}O}$ correlates well with $C_{12/2.2}$.  The color
temperatures are indicated on the upper x-axis.  The open rectangles
denote the six stars ($<$~M6) with H$_2$O.  The early M-type stars
without H$_2$O are overlapped by diamonds.
There is a systematic dependence of $C_\mathrm {12/2.2}$ on
variable types: Miras show the deepest H$_2$O absorption, as well as
reddest $C_\mathrm {12/2.2}$, and the SRs are second.  Ls are scattered
in a regime from no-H$_2$O to lower end of the H$_2$O absorption in Miras
(see also Fig.~\ref{h2o_nirs} for variable-type dependence).
}
\label{h2o_nirs_iras}
\end{figure}

In Fig.~\ref{h2o_nirs_iras}, we plot $I_\mathrm{H_{2}O}$ versus
the color defined by the NIRS 2.2~$\mu$m band flux
and IRAS 12~$\mu$m flux ($F_{12}$) as:
\begin{equation}
 C_{12/2.2} = \log (F_{12} / F_{2.2}).
\end{equation}
$C_{12/2.2}$ is a measure of the IR excess due to circumstellar dust,
and is roughly equivalent to $K-[12]$, which is an indicator
of mass-loss rate in Miras (Whitelock et al. \cite{whitelock}).
Fig.~\ref{h2o_nirs_iras} shows boundary at 
$C_{12/2.2} \approx -1.0$ 
between early M-type stars with H$_2$O and those without H$_2$O
(we regard all stars below the $+2\sigma$ line in Fig.~\ref{h2o_nirs}
as early M-type stars without H$_2$O).
Furthermore, there is a clear correlation between $I_\mathrm{H_{2}O}$ and 
$C_{12/2.2}$, which implies that the H$_2$O absorption is related to 
the circumstellar dust
emission.
However, as we show below, the H$_2$O molecules are not necessarily 
located in the circumstellar envelope.
%
%

\begin{figure}
\vspace{0cm}
\resizebox{\hsize}{!}{\includegraphics{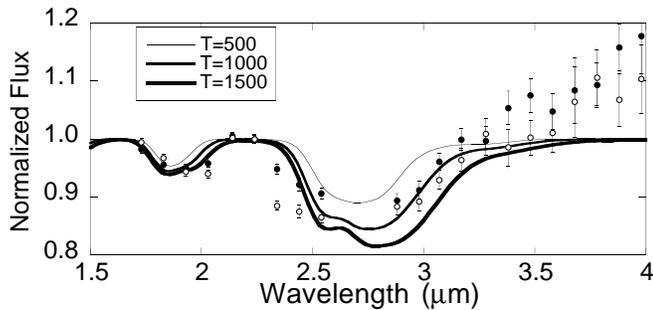}}
\caption{
The observed spectral profile of H$_2$O in \object{AK~Cap} (M2),
divided by the spectra of  \object{HR~1667} (M2; open circle) and
\object{HR~6306} (M2; filled circle),
and normalized to 1 at 2.2~$\mu$m. Both \object{HR~1667} and \object{HR~6306}
have no H$_2$O absorption lines and no dust emission.
The three lines indicate LTE model spectra
with a column density of 5$\times 10^{19}$~cm$^{-2}$,
and temperatures of 500, 1000, and 1500~K.
The excess at longer wavelength is possibly due to dust emission.
}
\label{h2o_akcap}
\end{figure}

 We estimate the excitation temperature ($T_\mathrm{ex}$), and the column
density ($N$) of H$_2$O molecules in early M-type stars.
The spectrum of a representative star \object{AK~Cap} is normalized with 
respect to the spectra of two early M-type stars, 
which have the same spectral types and similar near-IR color ($C_{2.2/1.7}$), 
but do not show H$_2$O absorption ($I_\mathrm{H_{2}O}$),
or dust excess ($C_{12/2.2}$).
The resulting normalized spectrum of \object{AK~Cap} is 
fitted by a simple plane parallel model with a uniform molecular layer
assuming local thermodynamic equilibrium (LTE)
(Fig.~\ref{h2o_akcap}). 
The H$_2$O line list is taken from Partridge \& Schwenke (\cite{partridge}).
The turbulent velocity is assumed to be 3~$\rm km~s^{-1}$.
For \object{AK~Cap}, we obtain a
reasonable fit for $T_\mathrm{ex} \approx 1000$--$1500~$K
and $N \approx 5\times10^{19}$~cm$^{-2}$.
A similar
analysis for \object{V~Hor} (M5III) results in $T_\mathrm{ex} \approx
1000$--$1500~$K and $N \approx 1\times10^{20}$~cm$^{-2}$. 
If we assume that the H$_2$O molecules were in the circumstellar envelope,
the mass-loss rate obtained from these column densities would exceed
$\rm 10^{-6}~M_{\sun}~yr^{-1}$
(assuming an abundance ratio of H$_2$O/H$_2$ $= 8\times10^{-4}$; Barlow et al. \cite{barlow}),
which is about a factor of 10--100 larger than those
expected for Ls and SRs
($\rm 10^{-7}$--$10^{-8}~\rm M_{\sun}~yr^{-1}$, Jura \& Kleinmann \cite{jurab}).
Thus, the H$_2$O molecules responsible for observed feature cannot be in
the circumstellar shell.

In contrast, our measured values for 
$T_\mathrm{ex}$ and $N$ are in good
agreement with results by Tsuji et al. (\cite{tsuji97}),
who suggested that the H$_2$O molecules in M-type stars 
are located in the layer above the photosphere.
Our numbers $T_\mathrm{ex}$ are also consistent with 
$T_\mathrm{ex} \approx 1150\pm200$ by Hinkle \& Barnes (\cite{hinkle})
for the `cool component' of the Mira variable \object{R~Leo},
which is an overlaying layer of photosphere.
Because the H$_2$O molecules responsible for the near-IR absorption
cannot be in the circumstellar shell, and because our results
are consistent with Tsuji et al. (\cite{tsuji97}) and
Hinkle \& Barnes (\cite{hinkle}), we conclude that they are
in an outer atmosphere, i.e. the layer above the photosphere, but
below the circumstellar envelope.

Numerical calculations of the atmospheres of
Miras (e.g. Bowen~\cite{bowen}; Bessell et al.~\cite{bessell96}) show
that the pulsation of the star extends the stellar atmosphere. 
Our observations show that such an extended region could also be
present in some SRs and Ls.
The dependence of H$_2$O intensity on variable types
as seen in Fig.~\ref{h2o_nirs} and \ref{h2o_nirs_iras} may result 
from differences in the physics of pulsation.
Not all Ls and SRs show H$_2$O absorption,
which is not surprising, because of their complexity 
(e.g. Jura \& Kleinmann \cite{jurab}).
Using the light curve of \object{V~Hor} (Mattei \cite{mattei}), we confirm
that \object{V~Hor} is probably an SR, although it shows 
a sudden increase of its visual magnitude
before the IRTS observation.
Unfortunately, no light curve is available for \object{AK~Cap}.

Hinkle \& Barnes (\cite{hinkle}) found that in Miras the H$_2$O in an outer
layer is responsible for the near-IR absorption, and we find 
the same situation in some
early M-type stars.  If we assume a constant abundance ratio of
H$_2$O$/$H$_2$, then $I_\mathrm{H_{2}O}$ is a measure of the total column
density in the outer layer.  Furthermore, $C_{12/2.2}$ is a measure
of the amount of the hot circumstellar dust, if the circumstellar shells of
the stars in Fig.~\ref{h2o_nirs_iras} have a similar dust composition.
Therefore, the total column density of the outer layer correlates with the
thickness of the circumstellar shell.  This may suggest that the
outer layer influences the mass loss of the star.\\

In conclusion, we demonstrate that H$_2$O absorption can be
seen in early M-type stars, and
that the H$_2$O molecules are located
in the outer atmosphere.
The observed correlation between the intensity of the H$_2$O absorption
and the mid-infrared excess implies that the extended atmosphere
is connected to the mass loss of the stars.

\begin{acknowledgements}
 The authors acknowledge Drs.~M.~Cohen and M.~Noda for their efforts on  
the NIRS calibration. 
M.M. thanks the Research Fellowships of the 
Japan Society for the Promotion of Science for the Young Scientists.
I.Y. acknowledges financial support from a NWO PIONIER grant.
M.M.F. thanks
Dr. H.A. Thronson at NASA Headquarters for discretionary funding, as well
as the Center of Excellency of the Japanese Ministry of Education.

\end{acknowledgements}

 
\end{document}